\documentclass[twocolumn]{aastex631}

\def\micron{$\mu$m}
\def\kms{km\,s$^{-1}$}

\def\s{$\sigma$}
\def\w{$\omega$}

\def\Msun{M$_\odot$}

\def\s{$\sigma$}

\def\l{$\lambda$}

\def\H2O{H$_{2}$O}
\def\C2H2{C$_{2}$H$_{2}$}
\def\CO2{$^{12}$CO$_{2}$}
\def\13CO{$^{13}$CO$_{2}$}
\def\13CO2{$^{13}$CO$_{2}$}

\begin{document}

\title{XUE. Molecular inventory in the inner region of an extremely irradiated Protoplanetary Disk}

\author[0000-0001-9698-4080]{Mar\'ia Claudia Ram\'irez-Tannus}
\affiliation{Max-Planck Institut f{\"u}r Astronomie (MPIA), K{\"o}nigstuhl 17, 69117 Heidelberg, Germany}

\author[0000-0001-8068-0891]{Arjan Bik}
\affiliation{Department of Astronomy, Stockholm University, AlbaNova University Center, 10691 Stockholm, Sweden}

\author{Lars Cuijpers}
\affiliation{Department of Astrophysics/IMAPP, Radboud University, PO Box 9010, 6500 GL Nijmegen, The Netherlands}

\author[0000-0002-5462-9387]{Rens Waters}
\affiliation{Department of Astrophysics/IMAPP, Radboud University, PO Box 9010, 6500 GL Nijmegen, The Netherlands}
\affiliation{SRON, Niels Bohrweg 2, Leiden, The Netherlands}

\author[0000-0003-4214-2464]{Christiane G{\"o}ppl}
\affiliation{Universit{\"a}ts-Sternwarte M{\"u}nchen,
Ludwig-Maximilians-Universit{\"a}t,
Scheinerstr.~1, 81679  M{\"u}nchen, Germany}

\author[0000-0002-1493-300X]{Thomas Henning}
\affiliation{Max-Planck Institut f{\"u}r Astronomie (MPIA), K{\"o}nigstuhl 17, 69117 Heidelberg, Germany}

\author[0000-0001-7455-5349]{Inga Kamp}
\affil{Kapteyn Astronomical Institute, Rijksuniversiteit Groningen, Postbus 800, 9700AV Groningen, The Netherlands}

\author[0000-0003-3130-7796]{Thomas Preibisch}
\affiliation{Universit{\"a}ts-Sternwarte M{\"u}nchen,
Ludwig-Maximilians-Universit{\"a}t,
Scheinerstr.~1, 81679  M{\"u}nchen, Germany}

\author[0000-0002-6137-8280]{Konstantin V. Getman}
\affiliation{Department of Astronomy \& Astrophysics, Pennsylvania State University, 525 Davey Laboratory, University Park, PA 16802, USA}

% START ALPHABETICAL ORDER HERE:

\author[0000-0003-0919-1512]{Germán Chaparro}
\affiliation{FACom, Instituto de Física - FCEN, Universidad de Antioquia, Calle 70 No. 52-21, Medellín 050010, Colombia}

\author[0000-0002-5398-8265]{Pablo Cuartas-Restrepo}
\affiliation{FACom, Instituto de Física - FCEN, Universidad de Antioquia, Calle 70 No. 52-21, Medellín 050010, Colombia}

\author[0000-0002-1198-3167]{Alex de Koter}
\affiliation{Anton Pannekoek Institute for Astronomy, University of Amsterdam, Science Park 904, 1098 XH Amsterdam, The Netherlands}
\affiliation{Institute of Astrophysics, Universiteit Leuven, Celestijnenlaan 200 D, 3001 Leuven, Belgium}

\author[0000-0002-5077-6734]{Eric D. Feigelson}
\affiliation{Department of Astronomy \& Astrophysics, Pennsylvania State University, 525 Davey Laboratory, University Park, PA 16802, USA}
\affiliation{Center for Exoplanets and Habitable Worlds, Pennsylvania State University, 525 Davey Laboratory, University Park, PA 16802, USA}

\author[0000-0002-4022-4899]{Sierra L. Grant}
\affiliation{Max-Planck Institut f{\"u}r Extraterrestrische Physik (MPE), Giessenbachstr. 1, D-85748, Garching Germany}

\author[0000-0002-9593-7618]{Thomas J. Haworth}
\affiliation{Astronomy Unit, School of Physics and Astronomy, Queen Mary University of London, London E1 4NS, UK}

\author{Sebastián Hernández}
\affiliation{FACom, Instituto de Física - FCEN, Universidad de Antioquia, Calle 70 No. 52-21, Medellín 050010, Colombia}

\author[0000-0002-0631-7514]{Michael A. Kuhn}
\affiliation{Centre for Astrophysics Research, University of Hertfordshire, Hatfield, AL10 9AB, UK}

\author[0000-0002-8545-6175]{Giulia Perotti}
\affiliation{Max-Planck Institut f{\"u}r Astronomie (MPIA), K{\"o}nigstuhl 17, 69117 Heidelberg, Germany}

\author[0000-0001-9062-3583]{Matthew S. Povich}
\affiliation{Department of Physics \& Astronomy, California State Polytechnic University, 3801 West Temple Ave, Pomona, CA 91768  USA}

\author[0000-0002-3887-6185]{Megan Reiter}
\affiliation{Department of Physics and Astronomy, Rice University, 6100 Main St - MS 108, Houston, TX 77005, USA}

\author[0000-0002-4650-594X]{Veronica Roccatagliata}\affiliation{INAF-Osservatorio Astrofisico di Arcetri, Largo E. Fermi 5, 50125 Firenze, Italy}
\affiliation{Department of Physics ``E. Fermi'', University of Pisa, Largo Bruno Pontecorvo 3, 56127 Pisa, Italy}
\affiliation{INFN, Sezione di Pisa, Largo Bruno Pontecorvo 3, 56127 Pisa, Italy}

\author[0000-0003-2954-7643]{Elena Sabbi}
\affiliation{Space Telescope Science Institute, Baltimore, MD 21218, USA}

\author[0000-0002-1103-3225]{Benoît Tabone}
\affiliation{Institut d'Astrophysique Spatiale, Universit\'e Paris-Saclay, CNRS,  B\^atiment 121, 91405 Orsay Cedex, France}

\author[0000-0002-7501-9801]{Andrew J. Winter}
\affiliation{Universit\'{e} C\^{o}te d'Azur, Observatoire de la C\^{o}te d'Azur, CNRS, Laboratoire Lagrange, 06300 Nice, France}
\affiliation{Universit\'{e} Grenoble Alpes, CNRS, IPAG, 38000 Grenoble, France}

\author[0000-0002-5456-523X]{Anna F. McLeod}
\affiliation{y, Department of Physics, Durham University, South Road,  Durham DH1 3LE, UK}
\affiliation{Institute for Computational Cosmology, Department of Physics, University of Durham, South Road, Durham DH1 3LE, UK}

\author[0000-0002-2190-3108]{Roy van Boekel}
\affiliation{Max-Planck Institut f{\"u}r Astronomie (MPIA), K{\"o}nigstuhl 17, 69117 Heidelberg, Germany}

\author[0000-0002-1284-5831]{Sierk E. van Terwisga}
\affiliation{Max-Planck Institut f{\"u}r Astronomie (MPIA), K{\"o}nigstuhl 17, 69117 Heidelberg, Germany}

\begin{abstract}

We present the first results of the eXtreme UV Environments (XUE) James Webb Space Telescope (JWST) program, that focuses on the characterization of planet forming disks in massive star forming regions. These regions are likely representative of the environment in which most planetary systems formed. Understanding the impact of environment on planet formation is critical in order to gain insights into the diversity of the observed exoplanet populations.  XUE targets 15 disks in three areas of NGC\,6357, which hosts numerous massive OB stars, among which some of the most massive stars in our Galaxy. Thanks to JWST we can, for the first time, study the effect of external irradiation on the inner ($< 10$\,au), terrestrial-planet forming regions of proto-planetary disks. In this study, we report on the detection of abundant water, CO, \CO2, HCN and \C2H2 in the inner few au of XUE 1, a highly irradiated disk in NGC\,6357. In addition, small, partially crystalline silicate dust is present at the disk surface. 
The derived column densities, the oxygen-dominated gas-phase chemistry, and the presence of silicate dust are surprisingly similar to those found in inner disks located in nearby, relatively isolated low-mass star-forming regions. Our findings imply that the inner regions of highly irradiated disks can retain similar physical and chemical conditions as disks in low-mass star-forming regions, thus broadening the range of environments with similar conditions for inner disk rocky planet formation to the most extreme star-forming regions in our Galaxy.  

\end{abstract}

\section{Introduction}\label{sec:intro}

The formation history of exoplanetary systems is constrained by comparing the demographics of the observed exoplanet population to the properties of planet-forming disks. Until the arrival of JWST, disks for which we were able to determine the composition as well as their physical structure (e.g. mass, radius) were all located in  nearby ($<500$\,pc) regions. 
However, observations \citep{2010ApJ...718..810S, 2016arXiv160501773G, 2018ApJ...860...77E, 2019A&A...628A..85V} and theory \citep{2008ApJ...675.1361F, 2020MNRAS.491..903W} converge on the idea that well-studied disks (in regions such as Taurus and Lupus), are not typical as most ($> 50$\%) stars and planetary systems form in massive star-forming regions \citep[e.g.][]{2019ARA&A..57..227K}  where they are exposed to intense UV radiation from nearby OB stars \citep{2022EPJP..137.1132W}. 

External photo-evaporation by far-UV photons \citep{2018MNRAS.478.2700W} is expected to dominate over stellar encounters \citep{2018MNRAS.475.5618B} in most environments  \citep[][]{2001MNRAS.325..449S, 2016arXiv160501773G}. External FUV fluxes $\gtrsim 10^3$\,G$_0$ can drive significant mass-loss rates from proto-planetary disks (PPDs). For instance, in the core of the Orion Nebula Cluster (ONC), at FUV fluxes of $\gtrsim 5 \times 10^4$\,G$_0$, mass-loss rates up to $\sim 10^{-6}$\,$\rm{M_\odot}~\rm{yr}^{-1}$ are possible \citep[][]{1998ApJ...499..758J,1999AJ....118.2350H,2004ApJ...611..360A,2016MNRAS.457.3593F,2018MNRAS.481..452H}. Due to these high mass-loss rates, the mass and dissipation time-scale of the gaseous disk are reduced  \citep{2004ApJ...611..360A,2019MNRAS.490.5678C,2020MNRAS.491..903W},
resulting in rapid outer disk dispersal. This inhibits the growth and inward drift of dust particles from the outer disk \citep{2020MNRAS.492.1279S, 2023MNRAS.522.1939Q} and the rate of pebble accretion needed for rapid growth from planetesimals to protoplanets. Rapid loss of gas and dust may curtail the growth of gas giants, resulting in entirely different planetary architectures.

ALMA studies show that the cold dust reservoir in the outer disks near massive stars in Orion is depleted by photo-evaporation \citep[][]{2014ApJ...784...82M,2017AJ....153..240A, 2019A&A...628A..85V, 2020A&A...640A..27V, 2023A&A...673L...2V}. Some inner disk studies of a large sample of young massive star clusters, combining X-ray ($Chandra$) and infrared ($Spitzer$) data,  have found correlations between disk lifetimes and UV flux, although this may depend on the mass or age of the region \citep[][]{2011ApJ...733..113R, 2012A&A...539A.119F, 2015ApJ...811...10R, 2016arXiv160501773G,2018MNRAS.477.5191R}. 

Several studies of disks around nearby solar-type stars with $Spitzer$ report on the detection of rotational and ro-vibrational emission of molecules such as water, CO, OH, \CO2, HCN and \C2H2 \citep[][]{2008Sci...319.1504C, 2010ApJ...720..887P, 2011ApJ...731..130S, 2013ApJ...766..134N, 2013ApJ...779..178P}. These molecules trace warm gas and are typically found to be confined to the inner 5 to 10\,au of PPDs \citep[e.g.][]{2009ApJ...701..142G, 2011ApJ...743..147N, 2018A&A...618A..57W}. 
Thanks to the increase of a factor of more than two orders of magnitude in sensitivity of JWST compared to \textit{Spitzer} we are able to extend these observations to larger distances ($\geq2$\,kpc). This allows us for the first time to study the physical properties and chemical composition of PPDs in extreme radiation regions.

With an age of 1--1.6~Myr \citep{2014ApJ...787..108G} and located at a distance of $\simeq 1.77$~kpc \citep[][based on Gaia DR2 data]{2019ApJ...870...32K}, NGC\,6357 is one of the youngest and closest massive star formation complexes, containing one of the most massive stars in our Galaxy \citep[Pis24-1, O4III(f+)+O3.5If*;][]{2003IAUS..212...13W} and more than 20 other O stars across the field. 
NGC\,6357 has the great advantage that it allows us to probe different radiation environments in each of its three sub-clusters \citep[Pismis\,24, G353.1+0.6, and G353.2+0.7;][]{2007ApJ...670..428C, 2012A&A...539A.119F, 2014ApJ...787..108G} which share the same age and distance but differ significantly in their FUV environment irradiating the PPDs \citep{ 2010A&A...515A..55R, 2020A&A...633A.155R}. 
We re-asses the distance to Pismis 24 using Gaia DR3 \citep{2022yCat.1355....0G} and find a distance of 1690~pc (see Appendix~\ref{sec:distance}).

\begin{figure*}[ht!]
\centering
\includegraphics[width=\hsize]{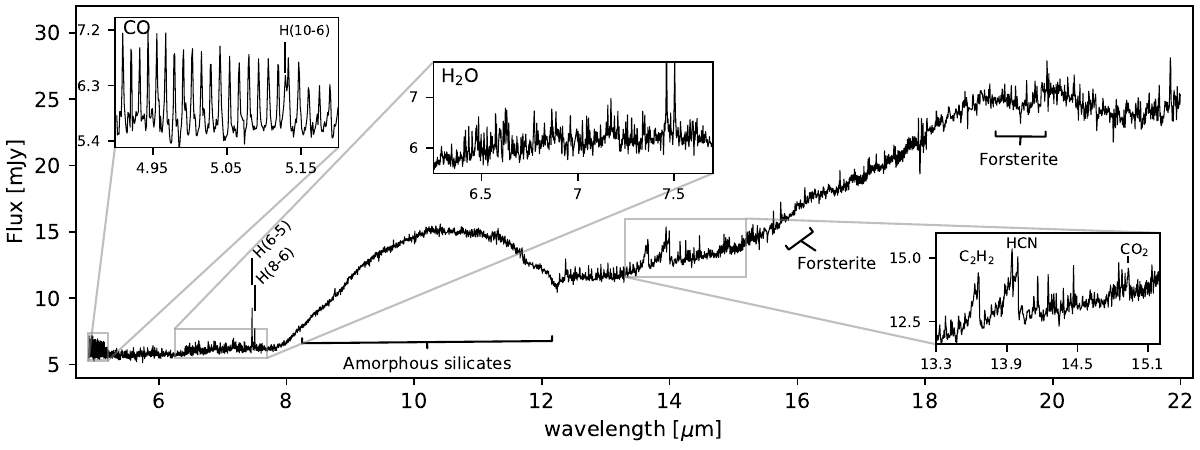}
  \caption{MIRI MRS spectrum of XUE\,1. The most prominent dust features are labeled. The insets show the P-branch transitions of the CO ro-vibrational fundamental band, the water emission around 7\,\micron\ and the 13 to 15\,\micron\ region featuring \C2H2, HCN, and \CO2. }
     \label{fig:MIRIspectrum}
\end{figure*}

The eXtreme UV Environments (XUE) program aims at characterizing the physical properties and chemical composition of 15 PPDs in NGC\,6357 with the Medium Resolution Spectrometer (MRS) of the Mid-InfraRed Instrument (MIRI).
In this paper we focus on the PPD XUE\,1 (17:24:40.098, $-$34:12:25.55). 
XUE\,1 is located in the sub-region Pis24 near the O-star binary Pis24-1. We infer an age of $\approx 0.7$~Myr for this source as the median age of stars in its vicinity from Table~6 of \citet{2022ApJ...935...43G}, based on the PARSEC 1.2S age scale.
Due to its location near several massive stars in NGC\,6357, we expect XUE\,1 to have been constantly exposed to a high radiation field throughout its life ($\sim10^5$\,G$_0$; Winter et al. $in\ prep.$). 
This unlike irradiated disks in the ONC, where a single O star ($\theta^1$~C) dominates the UV flux, and therefore PPDs migrate in and out of the proplyd regime ($\sim5\times10^4~G_0$) in short periods of time ($\sim0.1$~Myr).

In this paper we explore the impact of externally-driven photo-evaporation on PPDs located in an extreme FUV environment using MIRI MRS spectroscopy. 
We report on the chemical inventory of the inner few au of XUE\,1. 
By means of LTE and non-LTE slab models we determine the temperature, column density and emitting area of each of the molecules and we compare our findings with those for isolated, nearby-disks around stars of similar mass. 
In Section~\ref{sec:observations} we present the observational strategy and the data reduction. Section~\ref{sec:results} reports on the results and in Section~\ref{sec:discussion} we discuss and conclude this work.

\section{Observations} \label{sec:observations}

In this section we present the observations of XUE\,1. In Section~\ref{sec:NIRphotometry} we show the photometric observations of our target and in Section~\ref{sec:JWSTspectrum} we describe the MIRI MRS Observations.

\subsection{Near-infrared photometry}\label{sec:NIRphotometry}

XUE\,1 is part of a triple system (see Appendix~\ref{sec:counterpart}) where the third star (B) is at a distance of $1.5^{\prime\prime}$ ($\sim2600$\,au) from a binary system (A1+A2) whose components are $\approx 0.2^{\prime\prime}$ ($\sim300$\,au) apart. Based on the MIRI imaging we conclude that A1 is the star with the PPD giving rise to the observed MIRI spectrum (see Figure~\ref{fig:HST-MIRI-images} in Appendix~\ref{sec:counterpart}).

We derive the masses of XUE\,1~A1 and XUE\,1~A2 from their extinction corrected z-band magnitudes ($z_{A1} \simeq 17.9$~mag and $z_{A2} \simeq 18.9$~mag). We assume that both components are subject to the same extinction 
$A_V \sim 8.7$~mag as derived for the unresolved object A using UKIRT photometry \citep{2013ApJS..209...28K, 2013ApJS..209...32B}. The z-band is not much affected by the presence of accretion/disks, and assuming a reddening law of $A_z/A_V = 0.505$ \citep{1985ApJ...288..618R}, the comparison with the PARSEC 1.2S models \citep[][]{2014MNRAS.445.4287T, 2014MNRAS.444.2525C, 2015MNRAS.452.1068C} for the 0.7 Myr isochrone gives the masses $M_{A1} \approx 1.1$~M$_{\odot}$ and $M_{A2} \approx 0.6$~M$_{\odot}$ for a distance of 1690\,pc.

\subsection{JWST MIRI MRS Spectra}\label{sec:JWSTspectrum}

XUE\,1 was observed on August 3rd, 2022 as part of the eXtreme UV Environments (XUE) project in Cycle~1 \citep[GO-1759, ][]{2021jwst.prop.1759R} with the MIRI MRS \citep[][]{2015PASP..127..584R, 2015PASP..127..646W, 2015PASP..127..595W}. 
All three wavelength settings (SHORT, MEDIUM and LONG) are used for the observations. 
A four-point dither optimized for a point source was performed in the negative direction. The observations are obtained in the FASTR1 readout mode with 40 groups per integration and two integrations per dither position. This results in a total exposure time of 900 seconds per wavelength setting and 2700 seconds in total.
No dedicated background observations were taken for this observation. The background in the observation is dominated by the emission of the surrounding H$_{\rm II}$ region and PDR and spatially highly variable.

We reduced the data using the JWST Pipeline version 1.9.4, with context 1046 of the Calibration Reference Data System \citep[CRDS;][]{bushouse_howard_2023_7577320}.
The \texttt{Detector1} part of the pipeline was run without modifications. In \texttt{Spec2} we implemented a custom background subtraction; after the WCS assignment, we performed a background subtraction by removing the two nodding positions from each other. This minimizes the effect of the variable background with respect to taking a dedicated background observation. 
We then run the standard steps in \texttt{Spec2}, including the residual fringe correction. 
In \texttt{Spec3}, we skipped the outlier rejection step, as it causes artifacts in the extracted spectrum due to the undersampling of the PSF. The final cubes are combined using the \emph{drizzle} algorithm.

Both components A (A1+A2) and B are visible in the MRS data cubes but the two components in source A are not resolved.
To identify the disk-bearing source we used archival HST/ACS imaging from the Hubble Legacy Archive and the simultaneous MIRI F560W imaging corresponding to our program (see Appendix~\ref{sec:counterpart}). 
We measured the position of source A in the datacube and run the \texttt{extract1d} step with the position of the source as input. We extracted the source with a modified aperture 2 times smaller than the default value, linearly increasing from 0.3\arcsec\ at 5 micron to 1.28\arcsec\ at 22\micron. We use the aperture corrections corresponding to this extraction aperture. The full $4.9-22$\,\micron\ spectrum of XUE\,1 is shown in Figure~\ref{fig:MIRIspectrum}.

\section{Results} \label{sec:results}

The spectrum of XUE\,1 presented in Figure~\ref{fig:MIRIspectrum} shows a rising continuum from warm dust in the disk surface layers with amorphous silicate emission at 10\,\micron\  and 18\,\micron\, as well as forsterite features at 16.3\,\micron\ and 20\,\micron\ . The shape and strength of the dust features are typical for passive disks heated by stellar photons. In addition to the dust features, the MIRI spectrum displays a wealth of molecular emission from \H2O, CO, HCN, \C2H2, and \CO2.  In this section we first present the temperature derivation of the CO fundamental via the rotation diagram. Then we describe the results of the modeling procedure with slab models. The detailed spectral analysis leading to this section is described in Appendix~\ref{sec:analysis}.

\subsection{CO}\label{sec:results_CO}
\begin{figure}
\centering
\includegraphics[width=\hsize]{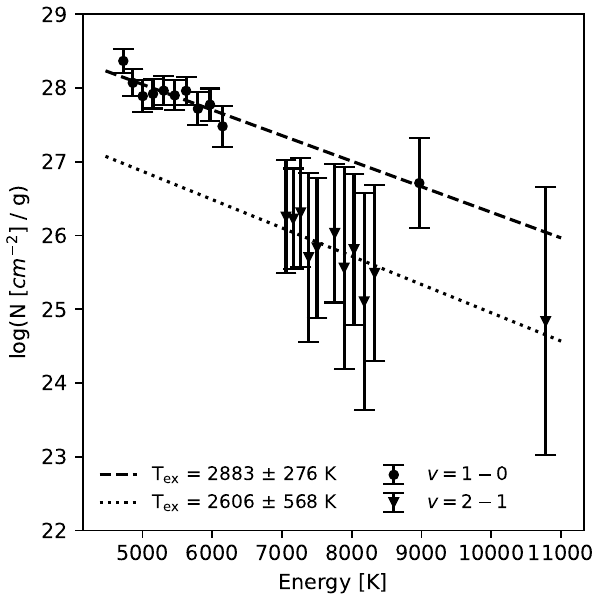}
  \caption{Rotation diagrams for the $v=1-0$ (dots), $v=2-1$ (triangles). The dashed and dotted lines show linear fits to each vibrational level.}
     \label{fig:rotationDiagram}
\end{figure}

We detect the $\mathcal{P}$-branch CO fundamental emission between 4.9 and 5.35 \micron\ with upper $J$ levels between 24 and 46 for the $v=1-0$ transitions and between 19 and 41 for the $v=2-1$ transitions. 
Using the rotational diagram method \citep{1999ApJ...517..209G} we estimate the excitation temperature and column density of the CO molecule. We fit the unblended lines with Gaussian profiles to find their flux (see Table~\ref{tab:CO_fluxes} in Appendix.~\ref{sec:CO_rotation} for the measured fluxes). 
We then calculate their specific intensity by assuming the emission originates from a disk annulus with a radius of 0.5\,au (see below) and a distance of 1690\,pc. 
The rotation diagrams in Figure~\ref{fig:rotationDiagram} are consistent with straight lines, suggesting LTE and optically thin conditions. 
We derive excitation temperatures of $T_{\rm ex}=2883\pm276\,{\rm K}$ for the  $v=1-0$ and of $T_{\rm ex}=2606\pm568\,{\rm K}$ for the $v=2-1$ bands.
A different choice of emitting area and/or distance would only affect the absolute values for the column densities, but does not change the temperature estimate nor the relative column density of $v=1-0$ with respect to $v=2-1$. 

The fact that we can perform linear fits for each vibrational level indicates that the rotational level populations within each vibrational level are in LTE. 
However, the measured column densities of the $v=2-1$ are lower than those of $v=1-0$, indicating that collision partner densities are not high enough to significantly populate the $v=2-1$ level, requiring non-LTE level populations (see also Appendix~\ref{sec:analysis}).

\begin{figure*}[ht!]
\centering
   \includegraphics[width=\hsize]{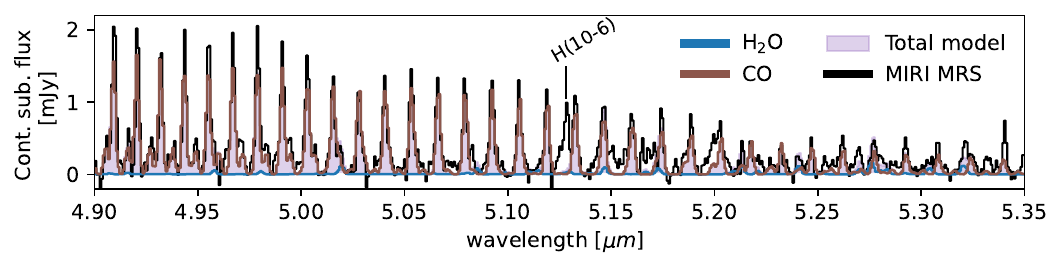}\\
   \includegraphics[width=\hsize]{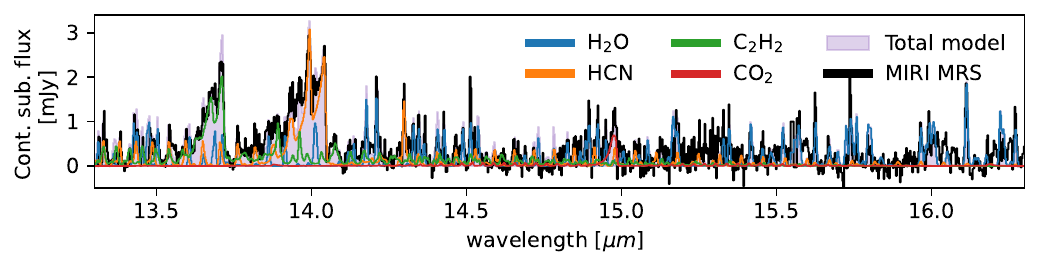}
\caption{
Continuum-subtracted MIRI spectrum of XUE~1 (black) with the best-fit slab models.
Molecules are shown with colors and the purple shaded area shows the total model spectrum. Top: Region between 4.9 and 5.35\,\micron\ including CO (brown) and \H2O (blue). Bottom: Region between 13 and 16\,\micron\ including \H2O (blue), \C2H2 (green), HCN (orange), and \CO2 (red).
}
\label{fig:fits}
\end{figure*}

We fit \texttt{radex} non-LTE slab models \citep[][]{2007A&A...468..627V} to the $v=1-0$ and $v=2-1$ simultaneously, including blended lines, following the methodology described in \citet{2023NatAs...7..805T} (see Section~\ref{sec:modeling_CO} for more details). The best fitting non-LTE model for CO is shown in the top panel of Figure~\ref{fig:fits} and the best fitting parameters are listed in Table~\ref{tab:molec_properties}. It has a temperature of 2300\,K, which is consistent with the results from the rotation diagrams. 
The best-fitting column density is $N=3\times10^{17}$\,cm$^{-2}$ and the emitting radius is $R=0.44$\,au. 
The density of collisional partners (see Appendix~\ref{sec:analysis}) of the best-fitting model is $n_{<H>}=6\times10^{12}$\,cm$^{-3}$ which is close to the critical density for $v=1-0$ \citep[][]{2013A&A...551A..49T}. 
This indicates that collisions populate the rotational levels within the v=0 according to LTE, but the densities are to low to bring the vibrational temperature to the gas temperature. 
The temperature is well constrained to values between 2000 and 2800\,K and $N$ is well constrained to values between $10^{16}$\,cm$^{-2}$ and $10^{18}$\,cm$^{-2}$ (see $\chi^2$ map in Appendix~\ref{app:chi2_maps}).
However, it has been shown that including the low-energy lines, not present in the MIRI spectrum, could lead to higher column densities and lower temperatures \citep[e.g.][]{2022AJ....163..174B}.
NIR ($\sim4.2 - 5$\,\micron) observations are needed to better constrain the CO fundamental's physical parameters.
In order to fit the CO lines to the slab models we need a lower resolution than that of the MRS at 5\micron\ \citep[][]{2023A&A...675A.111A}.
The CO lines are spectrally resolved with a deconvolved FWHM of 51\,\kms. 
Assuming that the emission originates in a disk annulus and Keplerian rotation, we can calculate the characteristic radius.
We adopt a mass of 1~\Msun\ and, as the inclination is not known, assume a value of 60$\degr$ for XUE\,1. This results in a characteristic radius of $0.25\pm0.09$\,au which is in agreement with the best fitting slab model. Assuming an edge-on disk would lead to a radius of 0.34\,au while an inclination of 30$\degr$ would result in a 0.09\,au radius.

\subsection{Spectral region between 6\,\micron\ and 17\,\micron}

We fit the spectra of water around 7\micron\ and in the 13-17\,\micron\ region with LTE slab models \citep[][see Appendix~\ref{sec:fit_others} for details]{2023NatAs...7..805T}. 
The bottom panel of Figure~\ref{fig:fits} shows the best-fit model to the spectral region between 13 and 16.3\,\micron\ and the best-fit model parameters for each molecule are listed in Table~\ref{tab:molec_properties}. The $\chi^2$ maps resulting from the fitting procedure for each molecule are shown in Appendix~\ref{app:chi2_maps}. From these maps it is possible to read the typical uncertainties on the best-fit parameters; the red, orange, and yellow lines show the 1, 2, and 3-$\sigma$ levels. 
In addition to the three free parameters ($T$, $N$, and $R$) we also list the total number of molecules ($\mathcal{N}_{\rm tot} = \pi N R^2$), which is well constrained for optically thin emission.

\begin{splitdeluxetable}{ccccccBcccc} \label{tab:molec_properties}
\caption{Best-fit model parameters for the studied molecules. Columns 2, 3, 4, 5, and 6 (top) show the disk properties obtained when leaving the radius as a free parameter. Columns 6, 7, 8 (bottom) show the $N$, $T$, and $<n>_{H}$ for a radius of 0.5\,au.} 
    \tablehead{
    \colhead{Species} & \colhead{$N$} & \colhead{$T$} & \colhead{$R$} & \colhead{$\mathcal{N}_{\rm tot}$} & \colhead{$<n>_{H}$} & \colhead{Species} & \colhead{$N_{\rm0.5\,au}$} & \colhead{$T_{\rm0.5\,au}$} \\
     \colhead{} & \colhead{[cm$^{-2}$]} & \colhead{[K]} & \colhead{[au]} & \colhead{[mol.]} & \colhead{[cm$^{-3}$]} & \colhead{[cm$^{-2}$]} & \colhead{[K]} \\ 
     }
    \startdata
        H$_2$O [7\,$\mu$m] & $2.2\times10^{18}$ & 975 & 0.13 & $2.6\times10^{43}$ & $-$ & H$_2$O [7\,$\mu$m] & $6.8\times10^{16}$ & 1150 \\ 
        H$_2$O [15\,$\mu$m] & $6.8\times10^{19}$ & 550 & 0.46 & $1.0\times10^{46}$ & $-$ & H$_2$O [15\,$\mu$m] & $2.2\times10^{19}$ & 575 \\ 
        HCN & $2.2\times10^{17}$ & 575 & 0.57 & $5.0\times10^{43}$ & $-$ & HCN & $4.6\times10^{16}$ & 675 \\ 
        C$_2$H$_2$ & $2.2\times10^{18}$ & 475 & 0.23 & $8.2\times10^{43}$ & $-$ & C$_2$H$_2$ & $3.2\times10^{17}$ & 550 & \\ 
        CO$_2$ & $2.2\times10^{14}$ & 450 & 5.30 & $4.3\times10^{42}$ & $-$ & CO$_2$ & $3.2\times10^{16}$ & 450 \\ 
        CO & $3.0\times10^{17}$ & 2300 & 0.44 & 4.1$\times10^{43}$ & $6.0\times10^{12}$ &  &  &  & \\
    \enddata
\end{splitdeluxetable}

\subsubsection{\H2O}

Many \H2O\ emission lines are present in the spectrum of XUE\,1, both in the region around 7\,\micron\, corresponding to the bending mode, and around 15\,\micron, corresponding to rotationally excited lines. For the fitting, we include emission of both ortho- and para-\H2O\ assuming a typical ratio ortho/para$=3$ \citep[e.g.][]{2012A&A...540A..84H, 2014A&A...572A..21M, 2016A&A...587A.139H, 2019A&A...632A...8P}. The region from 6 to 8\,\micron\ is best represented by a LTE slab model with $T=975$\,K, $N=2.2\times10^{18}$\,cm$^{-2}$ and an emitting area given by $R=0.13$\,au, whereas the 13--17\,\micron\ region is best reproduced by a model with $T=550$\,K, $N=6.8\times10^{19}$\,cm$^{-2}$ and an emitting area given by $R=0.46$\,au. For the hotter \H2O\ we can exclude temperatures $<300$\,K, but the upper limit on the temperature is poorly constrained. For the cooler \H2O\ component the temperature is better constrained to values between $\sim450$ and $\sim700$\,K. 
It has been pointed out in the literature that the shorter wavelengths probe \H2O\ at higher temperatures, while the longer wavelengths probe the colder regions \citep[][]{2016ApJ...818...22B, 2023AJ....165...72B, 2023arXiv230709301G}. For XUE\,1, although the best fitting values agree with this behavior, when taking into account the confidence intervals the temperature cascade does not seem to be as extreme, similarly as in PDS\,70 \citep{2023arXiv230712040P}.

\subsubsection{\CO2}

We detect the \CO2\ $\mathcal{Q}$ branch at 14.9\,\micron. The best-fitting model for \CO2\ has a temperature of 450\,K, a column density of $2.2\times10^{14}\,{\rm cm}^{-2}$ and an emitting area given by $R=5.3$\,au. Nevertheless, the column density is in the optically thin regime, so $N$ and $R$ are degenerate. 
The temperature is well constrained to values between 300 and 650\,K. 
We do not find evidence for the \CO2\ hot bands at 13.9 and 14.2\micron\ nor for the presence of CO$_2$ isotopologues as found by \citet{2023ApJ...947L...6G} around the low-mass T\,Tauri star GW\,Lup. This supports our findings that CO$_2$ is likely optically thin.

\subsubsection{\C2H2\ and HCN}

We detect the $\mathcal{Q}$ branches of \C2H2\ and HCN around 13.7 and 14\,\micron, respectively, including the HCN hot-band at 14.3\,\micron.
For HCN the temperature is well constrained for a range between $\sim550$ and $\sim650$\,K. For \C2H2\ the best-fitting models have temperatures between $\sim300$ to $\sim800$\,K. We find column densities of $2.2\times10^{17}\,{\rm cm}^{-2}$ and $2.2\times10^{18}$\,cm$^{-2}$ for HCN and \C2H2, respectively. 
For \C2H2\ the best fit model is in the optically thick regime, but models with higher temperatures and in the optically thin regime are not possible to exclude. 
From the best-fitting models we find emitting areas given by radii of $R=0.57$ and 0.23\,au.
After subtracting all the best-fitting models to the observed spectrum we do not find evidence for OH emission.

As \CO2\ and possibly \C2H2\ are in the optically thin regime, and therefore we cannot constrain $R$ and $N$ independently, we also calculate the best-fit parameters for an assumed radius of 0.5\,au, based on radiative transfer models \citep[][]{2019A&A...631A..81G}. 
The results of the latter are listed in the last three columns of Table~\ref{tab:molec_properties}.

\section{Discussion and conclusions} \label{sec:discussion}

In this paper we present the first JWST-MIRI spectrum of an extremely irradiated PPD located in a 1~Myr old massive star-forming cluster. We detect \H2O, CO, \CO2, HCN, and \C2H2 in its inner disk as well as cristalline silicates. 

\subsection{Comparison to PPDs in low-mass star-forming regions}
\begin{figure*}[ht]
\centering
\includegraphics[width=\hsize]{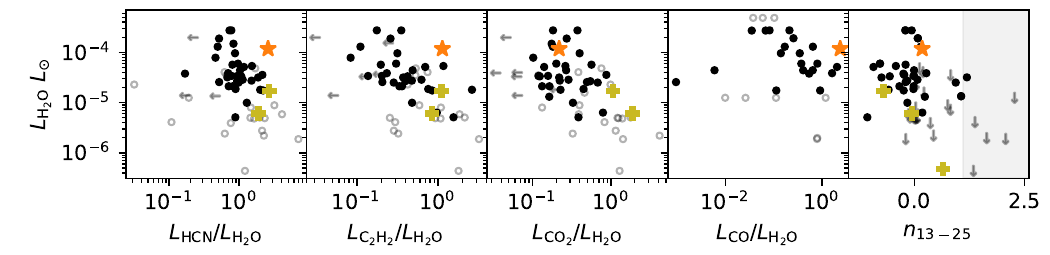}
  \caption{Comparison between the gas line luminosities and spectral index n$_{13-25}$ of XUE\,1 and the \textit{Spitzer} sample studied by \citet{2020ApJ...903..124B} and the CO emission from \citet{2022AJ....163..174B} (black dots). The y-axis shows the 17\,\micron\ \H2O\ luminosity. From left to right the x-axis show the ratio of HCN, \C2H2, CO2\ and CO to \H2O\ luminosities. The x-axis of the right-most panel shows the spectral index n$13-25$ where shaded region shows the location of disks with large gaps and/or cavities. The measurements for the JWST spectrum of XUE\,1 are shown with the orange stars. The upper limits from \citet{2020ApJ...903..124B} and \citet{2022AJ....163..174B} are indicated with grey arrows and the empty circles show the sources with upper limits both in the numerator and denominator. The yellow plus-signs show the location of the MIRI spectra published to date; GW\,Lup \citep{2023ApJ...947L...6G}, EX\,Lup \citep{2023arXiv230108770K} and PDS\,70 \citep{2023arXiv230712040P}.}
     \label{fig:Banzatti20_specind}
\end{figure*}

We compare our measurements with $Spitzer$ and newly obtained JWST observations of PPDs that are not strongly irradiated, i.e. in low-mass star-forming regions. 
Figure~\ref{fig:Banzatti20_specind} shows the line luminosity of \H2O, measured from the 17\,$\mu$m lines, against the HCN/\H2O, \C2H2/\H2O, CO/\H2O and \CO2/\H2O of XUE\,1 in comparison with a sample of T\,Tauri stars located in 
nearby ($<200$\,pc), young (1-3\,Myr) star-forming regions from  \citet{2020ApJ...903..124B,2022AJ....163..174B} and the newly obtained JWST observations of isolated PPDs \citep{2023arXiv230108770K, 2023ApJ...947L...6G, 2023arXiv230712040P}. We find that the H$_2$O luminosity and the \CO2/\H2O ratio are at the high end of the distribution. However, The HCN/\H2O,  \C2H2/\H2O and CO/\H2O ratios are higher than found in the T~Tauri sample. 

High line luminosities of molecular species are a signpost of a high energy input into the disk, which can be intrinsic, i.e. due to a high stellar luminosity and/or a high gas accretion rate,  or extrinsic, i.e. a strong external irradiation field, or a combination of these effects. Since the line luminosities of XUE\,1 are in the range of those observed in non-irradiated  disks (Figure~\ref{fig:Banzatti20}; but note the high \C2H2 and HCN line luminosities; see below), disk irradiation is not needed to explain our observations. 

\cite{2023arXiv230712040P} compare the H$_2$O luminosity of a sample of disks from \cite{2020ApJ...903..124B,2022AJ....163..174B} to the continuum spectral index n$_{13-30}$, which is a measure of the dust depletion of the inner disk. We follow this approach, but we calculate the n$_{13-25}$ from the \textit{Spitzer} spectra from the Spexodisks database \citep[][]{2010ApJ...720..887P, 2011ApJ...731..130S, 2020ApJ...903..124B} to allow direct comparison to our MIRI data. The right panel of Figure~\ref{fig:Banzatti20_specind} shows that XUE\,1 does not have a dust depleted inner disk, which implies that the inner disk dust and gas content XUE\,1 is comparable to the average T~Tauri disks in nearby low mass star forming regions.

In summary, we find no substantial evidence for an externally irradiated disk surface in XUE\,1. This is surprising because models of disk spectra for continuous, irradiated disks show much stronger line enhancements than detected for XUE\,1 \citep{2015A&A...582A.105A}. These strong line enhancements are the result of the entire disk surface being warm enough to contribute to the line emission. In the following we discuss two scenarios that can explain our observations, i.e. the disk is shielded from UV irradiation, or the disk is truncated.

\begin{figure}[ht!]
    \plotone{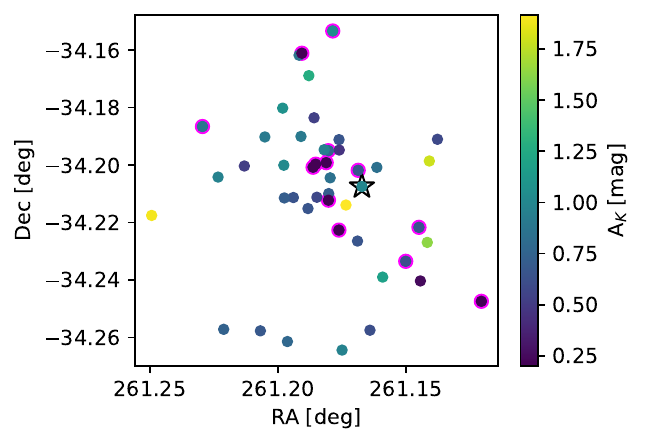} \\
    \plotone{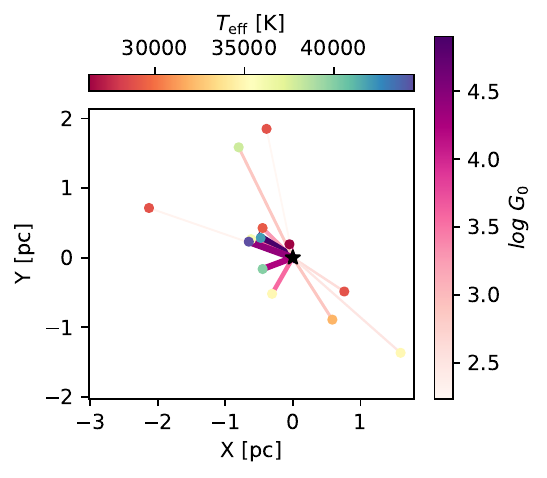}
    \caption{\textit{Top:} Extinction A$_K$ for a sample of stars in Pis24, A$_K$ is shown in colors. The position of XUE\,1 in this diagram is indicated with a star. The O stars from the bottom panel are indicated with magenta borders. \textit{Bottom:} Radiation field towards XUE\,1. The lines show the 2D distance from the massive stars to XUE\,1 (indicated with the black star) and the colors of the dots show their temperature. The FUV radiation felt by XUE\,1 from each massive star is shown by the color and width =of the lines}.
    \label{fig:ext_radiation}
\end{figure}

\subsection{Is XUE\,1 shielded from FUV radiation?}

A scenario in which XUE\,1 is shielded from the UV radiation from the massive stars could explain why its spectral properties in the MIRI spectral range resemble those of nearby disks. The physical and chemical properties of the inner-most parts of the disks traced by MIRI are dominated by the radiation from the central star. 
Therefore, if XUE\,1 is shielded from the radiation, given the mass of the central star, there would be no reason to expect the MIRI MRS spectrum to be different from those of nearby regions. 
In the top panel of Figure~\ref{fig:ext_radiation} we show the extinction A$_K$ reported by \citet{2020A&A...633A.155R} for a sample of stars in Pis24 including the extiction measured for XUE\,1 in this work. 
The extinction towards XUE\,1 does not seem higher that that of the neighboring stars. 
This is consistent with the lack of a significant dust cloud between XUE\,1 and the O-type stars in the Herschel maps. However, the resolution of Herschel is not high enough to detect small-scale molecular features.   

The bottom panel of Figure~\ref{fig:ext_radiation} shows the projected distribution of massive stars (T$_{\rm eff} \geq 24\,000$~K) in Pis24 with respect to the position of XUE\,1. The external FUV radiation felt by XUE\,1 assuming a 2D projected distance is represented by the colors and width of the lines. The stars contributing the most to the FUV radiation are all located towards the left side of the plot. Given this asymmetric FUV radiation field, it is not possible to discard the presence of a coherent dust structure, like a filament, shielding XUE\,1 from most of the external radiation. 

Nevertheless, \citet{2014ApJ...787..108G} show that that heavily embedded objects within clouds, such as XUE\,1 (A$_V\gtrsim10$~mag), may undergo an environmental transformation due to cloud dispersal. These objects can become less absorbed (A$_V\sim5$~mag), and thus be significantly influenced by the radiation emitted by nearby massive stars, on timescales of less than 0.5~Myr.  Therefore, it is still possible that XUE\,1 has been exposed to the UV radiation at some point during its evolution.
It is also possible that XUE\,1 and the massive stars are not exactly at the same distance. If that were the case, the FUV field felt by XUE\,1 could be substantially lower than reported in Figure~\ref{fig:ext_radiation}.

\subsection{Is XUE\,1's disk truncated?}

Disk truncation would result in a change of the SED. Given the wavelength coverage of the MIRI MRS spectrum and the lack of longer wavelength data, disks as small as a few au would all show a similar spectrum in the MIRI data, i.e. poorly constraining the outer disk radius. The line emission provides much stronger constraints on the disk outer radius. \citet{2015A&A...582A.105A} predict the mid-infrared emission of externally irradiated disks to be much stronger than in isolated disks. This because of an increase of the line emitting area.  If XUE\,1's disk were truncated, it would not be possible to form a hot layer of gas extending beyond the inner disk and therefore the emitting area would be similar to that observed in nearby disks, which is dominated by the radiation from the central star and by accretion heating. Therefore, a truncation of XUE\,1's disk could explain the fact that we do not see extraordinarily strong emission lines. 

The observed HCN and \C2H2 line luminosities of XUE\,1 are higher than in the Banzatti sample, while this is not the case for \CO2\ . We note that this is consistent with disk irradiation, because in non-irradiated disks the \C2H2\ and HCN emission is more concentrated in the innermost disk, while \CO2\ has a larger emitting area. In an irradiated disk \C2H2\ and HCN will therefore be relatively more enhanced. However, we note that other effects such as a higher inner disk C/O ratio (perhaps caused by a lower inward pebble flux) may also result in high \C2H2\ and HCN line luminosities. 

Disk truncation may also be traced by the presence or absence of PAHs. For PAHs to emit in their infrared vibrational resonances, they need to be exposed to a UV radiation field. T~Tauri stars in low mass star forming regions only rarely show PAH emission \citep{2006A&A...459..545G}. This is most likely due to a lack of a strong enough stellar UV radiation field. Indeed, In the hotter Herbig Ae/Be stars PAHs are much more frequently detected, particularly in sources with a large dust disk gap and/or inner disk dust depletion \citep{2010ApJ...718..558A}. The PAH emission in Herbig stars is dominated by the outer disk surface layers \citep[e.g.][]{2006Sci...314..621L,2023A&A...674A..57Y} and traces the spatial distribution of gas. If the disk of XUE\,1 is exposed to a radiation field G$_0$ as high as indicated in the color bar of figure \ref{fig:ext_radiation}, we expect the full gas disk surface to contribute to the PAH emission, because such radiation fields are similar or exceed those provided by the central stars of disks surrounding Herbig stars. 
However, the background-subtracted spectrum of XUE\,1 (figure \ref{fig:MIRIspectrum}) does not show any evidence for PAH emission. The central star is probably too cool to produce sufficient UV photons. Assuming a similar abundance of PAHs in the disk of XUE\,1 as in those surrounding Herbig stars, this suggests that any gaseous disk surrounding XUE\,1 cannot not be spatially very extended, unless the disk is shielded from UV radiation. 

With the data at hand it is neither possible to fully discard the extinction nor the disk truncation scenario. 
Nevertheless, given the evidence presented we have a slight preference for the truncation scenario. 
Dedicated ALMA observations are needed to prove or disprove the hypothesis that XUE\,1's disk is truncated. 
Additionally we are performing a parameter study using the radiation thermo-chemical disk model ProDiMo \citep[][]{2009A&A...501..383W, 2017A&A...607A..41K} in which we study the effect of disk truncation and PAH abundance on the spectra of irradiated PPDs. With this study we will be able to better understand the MIRI MRS observations (Hernández et al. $in~prep.$).

\subsection{Consequences for planetary system formation}
With this first observation of water and other molecules the inner, terrestrial planet forming regions of a disk in one of the most extreme environments in our Galaxy, we show that conditions for (terrestrial) planet formation, routinely found in disks in low mass star forming regions, can also occur in massive star forming regions.
The fact that dust grain growth happens fastest in the inner disk \citep[e.g.,][]{2012A&A...539A.148B} and that there have been substructures detected in disks that could be caused by planets in systems at 0.5~Myr old \citep[][]{2020Natur.586..228S, 2018ApJ...857...18S} leaves the possibility that planet formation has already happened or has gone a long way at the age of XUE\,1 ($\sim 1$~Myr). Therefore, the fact that XUE\,1's disk might be truncated does not necessarily prohibit planet formation.

However, it is reasonable to expect different masses and orbital periods for the XUE1's planets, compared to those that formed in lower UV environments. 
For example, if planets form by pebble accretion then the young planets would be fed by a stream of pebbles from the outer disk. If the outer disk is dissipated, the supply of material that helps the planets grow is reduced \citep[see e.g.,][]{2023MNRAS.522.1939Q}. Indeed, outer disk truncation due to UV radiation is well documented by ALMA observations \citep[][]{2017AJ....153..240A, 2019A&A...628A..85V, 2020A&A...640A..27V, 2023A&A...673L...2V},  and our observations could be the first evidence based on MIRI observations that the outer disk radius in such an extreme environment has been truncated. 

XUE\,1 shows that the conditions for terrestrial planet formation can also happen in extreme environments. Nevertheless, the remaining observations from the XUE program are crucial to establish the frequency with which this occurs.

\section{Acknowledgments}

MCRT acknowledges support by the German Aerospace Center (DLR) and the Federal Ministry for Economic Affairs and Energy (BMWi) through program 50OR2314 ‘Physics and Chemistry of Planet-forming disks in extreme environments’.
AB acknowledges support from the Swedish National Space Agency (2022-00154). 
The research of CG~and TP~was supported by the Excellence Cluster ORIGINS which is funded by the Deutsche Forschungsgemeinschaft (DFG, German Research Foundation) under Germany’s Excellence Strategy - EXC-2094 - 390783311. TJH acknowledges funding from a Royal Society Dorothy Hodgkin Fellowship and UKRI guaranteed funding grant (EP/Y024710/1). IK acknowledges support from grant TOP-1 614.001.751 from the Dutch Research Council (NWO) and
funding from H2020-MSCA-ITN-2019, grant no. 860470 (CHAMELEON). 
GP gratefully acknowledge support from the Max Planck Society.
VR acknowledges the support of the Italian National Institute of Astrophysics (INAF) through the INAF GTO Grant  “ERIS \& SHARK GTO data exploitation”.  
BT has received funding under the Horizon 2020 innovation framework program and Marie Sklodowska-Curie grant agreement No. 945298. The research of BT is also supported by the Programme National “Physique et Chimie du Milieu Interstellaire” (PCMI) of CNRS/INSU with INC/INP co-funded by CEA and CNES. AJW has received support from the European Research Council (ERC) under the European Union’s Horizon 2020 research and innovation programme (PROTOPLANETS, grant agreement No. 101002188).

This work is based [in part] on observations made with the NASA/ESA/CSA James Webb Space Telescope. The data were obtained from the Mikulski Archive for Space Telescopes at the Space Telescope Science Institute, which is operated by the Association of Universities for Research in Astronomy, Inc., under NASA contract NAS 5-03127 for JWST. These observations are associated with program \#1759 and the specific observations analyzed can be accessed via \dataset[https://doi.org/10.17909/6nwv-6f56]{https://doi.org/10.17909/6nwv-6f56}.

\vspace{5mm}
\facility{JWST (MIRI MRS)}

\software{Astropy \citep{astropy:2013, astropy:2018, astropy:2022}, NumPy \citep{2020Natur.585..357H}, Matplotlib \citep{2007CSE.....9...90H}}

\appendix

\section{Optical/near-infrared counterpart of XUE~1} \label{sec:counterpart}

\begin{figure}[ht!]
\centering
    \includegraphics[width=\hsize]{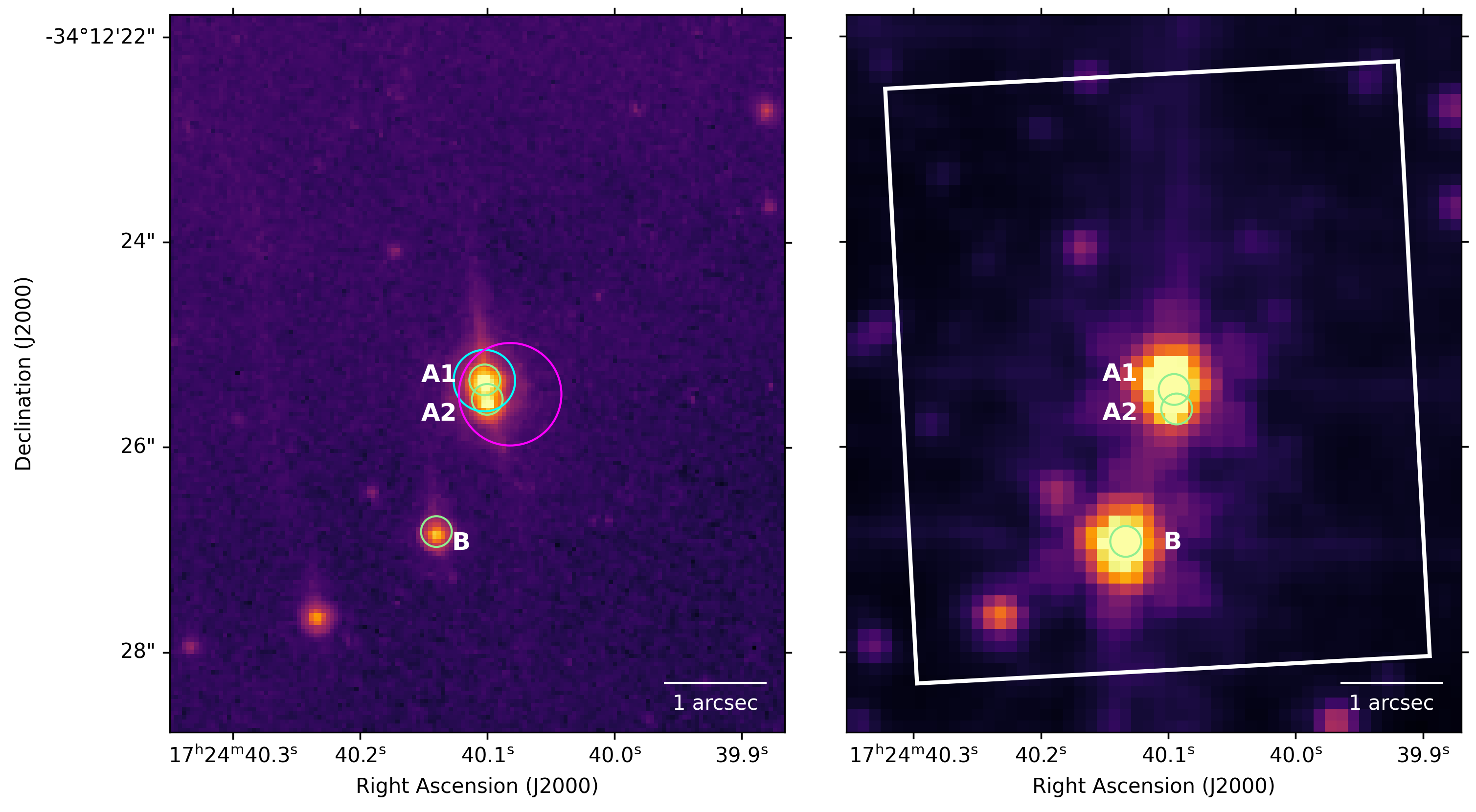} 
    \caption{\textit{Left:} HST/ACS  F850LP band image of the target position for XUE~1.  
   The three point-like objects are marked by green circles
with radii of $0.1\arcsec$.
The cyan circle on A1  marks the position of the Gaia DR3 source 5976051168205228416.
The magenta circle ($0.5\arcsec$ radius) marks the position of the
\textit{Chandra} X-ray source. A grid of J2000 coordinates is shown.
\textit{Right:} MIRI F560W image (log intensity scale) of the target position for XUE~1. The white box marks the observed field of view with MRS. The optical positions of the three point-like objects A1, A2, and B are marked by green circles
with radii of $0.1\arcsec$.}
    \label{fig:HST-MIRI-images}
\end{figure}

The original target of the MIRI observation was the 2MASS source
J17244012-3412263 ($J=14.77$, $H=12.34$, $K=11.54$).

Inspection of the MIRI data revealed that there are actually two mid-infrared sources near this position, with an angular separation of about $1.5\arcsec$.
The multiple nature of the source was confirmed by inspection of near-infrared images from the
''VISTA VVV: ZYJHKs Catalogue in the Via Lactea, Release 3'',
where the northern component (A)  has the source number 
vvv~J172440.11$-$341225.39 ($J=15.07$, $H=12.99$, $K=12.80$). The UKIRT magnitudes for this source are $J=14.93$, $H=13.53$, $K=12.59$ \citep[][]{2013ApJS..209...28K, 2013ApJS..209...32B}. 
The southern component (B) has the source number vvv~J172440.14$-$341226.70 ($J=15.14$, $K=11.78$).

The mid-infrared source XUE~1 coincides very well with component A.

 However, our inspection of the available optical HST/ACS archive
 images showed that the
northern component A is again resolved into two point-like components (denoted as A1 and A2), with an angular separation of just about $0.2\arcsec$.

The left panel of Fig.~\ref{fig:HST-MIRI-images} 
shows the HST ACS  F850LP image of XUE~1.

The J2000 coordinates of the three components, as measured in this HST  image, are $(\alpha, \delta) = $ (17:24:40.100, $-$34:12:25.36) for A1,
(17:24:40.098, $-$34:12:25.55) for A2,
and (17:24:40.138, $-$34:12:26.84) for B.

Component A1 coincides very well with the Gaia DR3
object 5976051168205228416, which has
magnitudes $G = 19.99$, $G_{BP} = 21.66$, and $G_{RP} = 18.43$; unfortunately,
 the listed parallax of $\varpi = (-0.66 \pm 0.69)$~mas does
not provide useful distance information.

Comparison of the HST and MIRI image coordinates of several point-like sources around the location of XUE~1, revealed a small discrepancy
in the astrometry ($\simeq 0.15\arcsec$ in RA and $\simeq 0.1\arcsec$ in Dec). After correcting for this small shift, we find that the strongest emission in
the MIRI image (right panel of Fig.~\ref{fig:HST-MIRI-images}) seems to be associated to source A1.

In order to estimate optical fluxes of the individual sources, we performed simple aperture photometry in the HST ACS images. 
In the F435W filter image, only component A1 is visible, and we estimated a flux of $F_\lambda \approx 3.1 \cdot 10^{-19}$~erg/cm$^2$/s/\AA, (roughly corresponding to a 
magnitude of $B \approx 24.4$).

In the F550M image, we estimate fluxes of 
$F_\lambda \approx 3.21\cdot 10^{-17}$~erg/cm$^2$/s/\AA\, ($\widehat{=}  \; V \approx 22.7$) for component A1, 
and $F_\lambda \approx 7.09\cdot 10^{-19}$~erg/cm$^2$/s/\AA\, ($\widehat{=} \; V \approx 24.3$) for component A2. 
In the F850LP image, we estimate fluxes of 
$F_\lambda \approx 5.73\cdot 10^{-17}$~erg/cm$^2$/s/\AA\, ($\widehat{=}  \; Z \approx 17.9$) for component A1, 
$F_\lambda \approx 2.28\cdot 10^{-17}$~erg/cm$^2$/s/\AA\, ($\widehat{=} \; Z \approx 18.9$) for component A2, 
and $F_\lambda \approx 8.71\cdot 10^{-18}$~erg/cm$^2$/s/\AA\, ($ \widehat{=} \; Z \approx 19.9$) for component B.

Finally, we note that XUE~1 was also detected as an  X-ray source in a \textit{Chandra} observation \citep{2019ApJS..244...28T}. The inferred X-ray column density $\log(N_H) = 22.2$~cm$^{-2}$ and X-ray luminosity $\log(L_X) = 30.4$~erg~s$^{-1}$ are consistent with the source extinction of $A_V \sim 9$~mag and mass of $M \sim 1$~M$_{\odot}$ for the more massive A1 component. This value that is a quite typical value for a young, approximately solar-mass star.

\section{Distance of XUE~1}  \label{sec:distance}

XUE~1 is located near the center of the stellar cluster Pismis~24.
Since no parallax information is available for XUE~1, 
we derived estimates of the mean distance to the Pismis~24 cluster in the following way: 

For an initial, rough, and preliminary estimate of the distance to the NGC~6357 region, we compiled a list of 61 early-type stars (spectral types O and B0--B3) from the literature \citep{2020A&A...642A..21R,2014ApJS..213....1T,2013ApJS..209...32B,2017A&A...607A..86R,2020A&A...633A.155R}, identified their Gaia DR3 \citep{2016A&A...595A...1G, 2023A&A...674A...1G} counterparts, corrected their parallaxes according to \citet{2021A&A...649A...4L}, and computed the 
 Maximum Likelihood 
estimate for the mean distance of this sample. This yielded an initial estimate of $ \langle D \rangle({\rm NGC 6357}) = (1.663 \pm 0.009)$~kpc; this value was used as prior information in the following steps of our distance determination. 

Next, we defined the Pismis~24 cluster region as a  $4.5^{\prime}$ radius circle centered on the apparent center of Pismis~24 at 
$(\alpha, \delta)[J2000]=(17^{\rm h}\, 24^{\rm m}\, 43.0^{\rm s}, -34^\circ\, 12^{\prime}\, 23^{\prime\prime})$. This region contains 21 of the O--B3 stars. 
We also collected a sample of X-ray selected objects in the same region from 
 \citet{2019ApJS..244...28T} and could identify 
 Gaia DR3 counterparts for 376 of these.

In the next step, we checked for the presence of possible 
fore-and background stars in our two samples, and 
excluded all stars for which the 3$\sigma$ distance interval 
(i.e.,~$[\,1 / (\varpi + 3 \sigma_\varpi)\: , \; 1 / (\varpi - 3 \sigma_\varpi)\,]$) lies outside the distance interval of $[1.663 \pm 0.1]$~kpc. 
This led to exclusion of one foreground star (2MASS J17244754-3415069, B1 V, $1/\varpi = 345^{+28}_{-24}$~pc) among the O--B3 stars,
and  13 foreground and 8 background objects among the 
X-ray selected objects. 

Finally, the mean distance to Pismis~24 was determined with the Bayesian inference code \textit{Kalkayotl} by \citet{2020A&A...644A...7O}, using a Gaussian prior distance of $(1.663 \pm 0.1)$~kpc. 
For the O--B3 stars this yielded $\langle D_{\rm OB3} \rangle = 1.687$~kpc with a central 68.3\% quantile (1$\sigma $ distance interval) of $[\,1.648 \;, \,1.726\,]$~kpc, and for the X-ray selected sample 
we found $\langle D_{\rm X}\rangle = 1.694$~kpc with a central 68.3\% quantile of $[\,1.656 \;,\, 1.732\,]$~kpc. 
Since these two distance estimates for the OB and the X-ray selected stars in Pismis~24 are very well consistent, we use $D \simeq  1.69$~kpc as the best estimate for the mean distance of Pismis~24, and thus for XUE~1.

\section{Spectral analysis} \label{sec:analysis}

In the following we describe the analysis procedure to derive the properties of the molecular species identified in the spectrum of XUE\,1.

\subsection{CO fundamental}

The MIRI spectrum contains a part of the $\mathcal{P}$-branch of the CO fundamental. Both the $v=1-0$ and the $v=2-1$ ro-vibrational transitions are detected in the spectra. In this section we present a first approach at deriving the excitation temperatures via the population diagram method \citep[][]{1999ApJ...517..209G} and we determine the temperature, column density and emitting area by means of fitting non-LTE slab models to the spectra.

\subsubsection{Rotation diagram}\label{sec:CO_rotation}

We selected the lines to be fitted by visually inspecting the spectrum and picking only those lines in which the $v=1-0$ and $v=2-1$ transitions are resolved and not contaminated by \H2O. We fitted the $v=1-0$ corresponding to upper $J$ levels between 24 and 33 and $J_{\rm up}=46$ and $v=2-1$ with upper $J$ levels between 18 and 28 excluding $J_{\rm up}=23$ and including $J_{\rm up}=41$. We determined the flux of each line as well as the continuum level by simultaneously fitting a zero-degree polynomial and all the selected lines with Gaussian profiles. To do the Gaussian fits we assumed an error of $10^{-4}$\,Jy, which corresponds to the typical error of line free regions. The errors in the line fluxes (reported in Table~\ref{tab:CO_fluxes}) were calculated as the square root of the covariance matrix. The specific intensity, $I$, was determined by integrating over each profile and dividing the observed flux by the solid angle, $\Omega$ assuming an emitting area of 0.5\,au and the distance to XUE1 to be 1690~pc. The column density was then calculated using the formula:

\begin{equation}
    N_j = \frac{4\pi \lambda_j I_j}{A_j h c}
\end{equation}

Where $\lambda$ is the central wavelength of each transition, $A_j$ is the Einstein A coefficient of each transition, $h$ is the Planck constant and $c$ the speed of light. Assuming that CO is in LTE, the energy levels can be described by a Boltzmann distribution. We can therefore express the relative column densities of any two excitation levels in terms of an excitation temperature $T_{\sc ex}$ as:

\begin{equation}
    \frac{N_i}{N_j} = \frac{g_i}{g_j}exp\left(-\frac{E_i - E_j}{kT_{\rm ex}}\right)
\end{equation}

where $g_j$ is the statistical weight of the rotational level, $E_j$ the energy of the upper level and $k$ is the Boltzmann constant. The excitation temperature $T_{\rm ex}$ correspond to the inverse slope of the $ln(N/g)$ vs $E_{\rm up}$ relation. 

\subsubsection{non-LTE slab models for CO}\label{sec:modeling_CO}

The ratio between the CO $v=1-0$ and the $v=2-1$ emission lines in XUE\,1 is particularly high and cannot be reproduced by LTE slab models. We therefore opt to fit the observed CO spectra with \texttt{radex} non-LTE slab models \citep[][]{2007A&A...468..627V} 
using the molecular data from \citet{2015ApJ...813...96S}. 
We follow the methodology from \citet{2023NatAs...7..805T}; We assume the line profiles to be Gaussian with a broadening $\Delta V = 4.7$~\kms as in \citet{2011ApJ...731..130S}. 
In the non-LTE models the density of collision partners n$_{<H>}$ is a free parameter, allowing us to decrease it in order to match the line ratio between the fluxes of the $v=1-0$ to the $v=2-1$ lines. 
The collision partners are H$_2$, He (0.1\,n$_{<H>}$), and electrons ($10^{-5}$\,n$_{<H>}$). 
Other free parameters are the column density ($N_{\rm CO}$), the excitation temperature ($T_{\rm ex}$), and the emitting area given by $\pi R_{\rm disk}^2$. Where $R_{\rm disk}$ does not necessarily correspond to the disk radius, but could also represent a ring with the same area.
Further studies are needed to account for the effect of fluorescence, which can also affect the $v=1-0$ to $v=2-1$ line ratio, but this is beyond the scope of this paper.

We fit the CO $v=1-0$ upper $J$ levels between 24 and 41 and $J_{\rm up}=46$ and $v=2-1$ with upper $J$ levels between 18 and 36, excluding $J_{\rm up}=23$, where the lines with $J_{\rm up}$ between 29 and 36 are blended with the $v=1-0$ lines with $J_{\rm up}$ between 34 and 41 and $J_{\rm up} = 41$. 
We use a $\chi^2$ minimization to find the best model that adjust to the observations. We compared the fluxes derived in Sect~\ref{sec:CO_rotation} with their respective errors to the model fluxes.

In an optically thin regime, $N_{\rm CO}$ and $R_{\rm disk}$ would be degenerate. Therefore, we can take advantage of the resolution of JWST in order to get an extra constraint on the emitting radius. We measure the width of the individual transitions and assume that the extra line broadening is due to Keplerian rotation. The projected rotational velocity can be expressed as:  

\begin{equation} \label{eq:vsini}
    v\sin{i} = c\frac{FWHM_{rot}}{2\lambda}
\end{equation}

where $i$ is the angle at which we observe the disk and $FWHM_{rot} = \sqrt{FWHM_{obs}^{2} - FWHM_{inst}^{2}}$. Where $FWHM_{inst}$ is the expected line width due to the resolution of the instrument \citep[$R\sim4000$;][]{2023A&A...675A.111A} and $FWHM_{obs}$ is the measured line width. The resulting velocity can then be used to provide additional constraints on the emitting surface radius provided we know the mass of the central star. Allowing us to circumvent the degeneracy between the column density and the emitting surface radius.

\subsection{LTE modeling of \H2O, HCN, \C2H2, and \CO2}\label{sec:fit_others}

In order to characterize the remaining molecular emission features present in the spectrum of XUE~1 we model them with the local thermal equilibrium (LTE) model presented in \citet{2023NatAs...7..805T} which has also been used in \citet{2023arXiv230712040P}, \citet{2023ApJ...947L...6G} and \citet{2023arXiv230709301G}. 
We included emission from \H2O (around 7 and 15\micron), \CO2, \13CO2, OH, \C2H2\ and HCN with molecular data from the HITRAN database \citep{2022JQSRT.27707949G}. 
As with CO Gaussian intrinsic line profiles with $\Delta V = 4.7$~\kms are assumed. For the HCN and \CO2\ $\mathcal{Q}$-branches, we included mutual shielding
from adjacent lines as described in \citet{2023NatAs...7..805T}. 
Under the LTE assumption we are able to fit the data with three free parameters: the gas temperature $T$, the column density $N$, and the emitting area as parameterized by an emitting radius $R_{\rm disk}$ ($\pi R_{\rm disk}^2$). As in the case of CO, $R_{\rm disk}$ is not necessarily the disk radius, but could also represent a ring.

In order to fit the MIRI-MRS spectrum, we first subtract the continuum locally by selecting line-free regions (see Table~\ref{tab:param_ranges}). We fit a spline function to determine the continuum level and subtract this from the MIRI spectrum. We then proceed to fit individual regions of the spectrum containing different molecules following a similar procedure to that described in \citet{2023ApJ...947L...6G}. 

For each molecule, we ran a grid of models varying $N$ in steps of 0.166 in $\log_{10}$-space, $T$ in steps of 25\,K in linear space, and $R_{\rm disk}$ from 0.01 to 10\,au in steps of 0.03\,au in $\log_{10}$-space. 
The best-fitting models were found by means of a $\chi^2$ fitting procedure. 
We measured the typical error in the flux in several line-free regions of the spectrum and adopted an error of $10^{-4}$\,Jy for the wavelength range between 13 and 17\,\micron.
We convolved the model spectra with a resolution $R=3000$ for the water around 7\micron\ and with $R=2500$ between 13 and 17\,\micron\ to mimic the limited spectral resolution of MIRI MRS. We then re-sampled the convolved model to the wavelength grid of the observations. 
The $\chi^2$ was calculated at given spectral regions in order to avoid regions of the spectrum that are strongly contaminated by other species while still including lines that are sensitive to temperature and optically thin lines. 
In the case of \H2O\ we selected the regions where to calculate $\chi^2$ by running an LTE model with $T = 800$\,K, $N=10^{17}$\,cm$^{-2}$ and $R=1$\,au to identify the spectral region where strong water lines are expected around 7\,\micron\ and 15\,\micron. The parameter ranges, as well as the wavelength regions used to calculate the $\chi^2$ are listed in Table~\ref{tab:param_ranges}.

\begin{table}[h]
\footnotesize
    \centering
    \renewcommand{\arraystretch}{1.1}
    \caption{Overview of the parameter ranges for the modeling procedure.}
    \begin{tabular}{cccc}
    \hline
    \hline
    Species & $\log(N)$ & $T$ & Fitting ranges \\
      & [cm$^{-2}$] & [K] & [\micron] \\
     \hline
    \H2O @7\,\micron & $15 - 22$ & $100 - 1400$ & $6.404 - 6.422$ \\
     &  &  & $6.445 - 6.456$\\
     &  &  & $6.467 - 6.478$\\
     &  &  & $6.630 - 6.641$\\
     &  &  & $6.678 - 6.689$\\
     &  &  & $6.781 - 6.792$\\
     &  &  & $6.858 - 6.871$\\
     &  &  & $7.039 - 7.051$\\
     &  &  & $7.202 - 7.212$\\
     \hline
    \H2O @15\,\micron & $15 - 22$ & $100 - 1400$ & $13.498 - 13.507$\\
     &  &  & $14.173 - 14.182$\\
     &  &  & $14.205 - 14.215$\\
     &  &  & $14.322 - 14.332$\\
     &  &  & $14.341 - 14.350$\\
     &  &  & $14.402 - 14.412$\\
     &  &  & $14.422 - 14.432$\\
     &  &  & $14.482 - 14.493$\\
     &  &  & $14.820 - 14.830$\\
     &  &  & $14.880 - 14.899$\\
     &  &  & $15.050 - 15.060$\\
     &  &  & $15.178 - 15.187$\\
     &  &  & $15.324 - 15.332$\\
     &  &  & $15.341 - 15.353$\\
     &  &  & $15.412 - 15.423$\\
     &  &  & $15.450 - 15.460$\\
     &  &  & $15.564 - 15.574$\\
     &  &  & $15.621 - 15.631$\\
     &  &  & $15.720 - 15.763$\\
     &  &  & $15.790 - 15.808$\\
     &  &  & $15.962 - 16.015$\\
     &  &  & $16.105 - 16.120$\\
     &  &  & $16.217 - 16.227$\\
     &  &  & $16.271 - 16.274$\\
     &  &  & $16.317 - 16.327$\\
    \hline
    HCN & $14 - 19$ & $200 - 1000$ & $13.980 - 13.997$\\
     &  &  & $14.022 - 14.057$\\
     &  &  & $14.290 - 14.308$\\
    \hline
    \C2H2 & $17 - 20$ & $100 - 1200$ & $13.562 - 13.748$ \\
    \hline
    \CO2 & $13 - 19$ & $100 - 1400$ &  $14.932 - 14.940$ \\
     &  &  & $14.959 - 14.990$\\
    \hline
    CO & $14 - 20$ & $100 - 3400$ & $4.900-5.152$\footnote{See Section~\ref{sec:modeling_CO} for the list of lines used.}\\
     &  &  & $5.18-5.19$\\
    \hline
    \end{tabular}
    \label{tab:param_ranges}
\end{table}

To further reduce the contamination by other species we follow the iterative procedure described in \citet{2023ApJ...947L...6G}. In short, we first fit \H2O, subtract the best-fitting model form the continuum-subtracted data and then proceed to do the same for HCN, \C2H2, and \CO2. We then subtract the composite best model of HCN, \C2H2 and \CO2 from the data and repeat the procedure for \H2O. We perform this process once and find no change in the best parameters nor any improvement on the residuals. 

%-------------------------------------------------------------------

\section{CO line fluxes for rotation diagram}
\label{sec:CO_fluxes}

\begin{table}[h]
\footnotesize
    \centering
    \renewcommand{\arraystretch}{1.1}
    \caption{Measured fluxes of the CO fundamental lines used for the rotation diagram and the CO $\chi^2$ modeling  (Section~\ref{sec:results_CO}). The blended lines (excluded from the rotation diagram) are indicated by listing both $v_{\rm up}$ and $J_{\rm up}$ levels.}
    \begin{tabular}{cccc}
    \hline
    \hline
    Wavelength & $v_{\rm up}$ & $J_{\rm up}$ & Flux \\
    \micron\ &  &  & ${\rm 10^{-16} erg\,s^{-1}\,cm^{-2} }$ \\
    \hline
    5.1887 & 1 & 46 & $1.853 \pm 1.135$ \\
    5.1192 & 1,2 & 41,36 & $3.158 \pm 0.889$ \\
    5.1057 & 1,2 & 40,35 & $3.551 \pm 0.933$ \\
    5.0924 & 1,2 & 39,34 & $3.814 \pm 0.945$ \\
    5.0792 & 1,2 & 38,33 & $3.443 \pm 0.91$ \\
    5.0661 & 1,2 & 37,32 & $3.942 \pm 0.96$ \\
    5.0532 & 1,2 & 36,31 & $3.871 \pm 0.919$ \\
    5.0405 & 1,2 & 35,30 & $3.907 \pm 0.989$ \\
    5.0279 & 1,2 & 34,29 & $2.971 \pm 0.919$ \\
    5.0154 & 1 & 33 & $3.286 \pm 0.911$ \\
    5.0031 & 1 & 32 & $4.318 \pm 0.945$ \\
    4.9908 & 1 & 31 & $3.998 \pm 0.907$ \\
    4.9788 & 1 & 30 & $4.979 \pm 0.933$ \\
    4.9668 & 1 & 29 & $4.602 \pm 0.933$ \\
    4.9550 & 1 & 28 & $4.785 \pm 0.925$ \\
    4.9434 & 1 & 27 & $4.461 \pm 0.884$ \\
    4.9318 & 1 & 26 & $4.205 \pm 0.904$ \\
    4.9204 & 1 & 25 & $4.89 \pm 0.901$ \\
    4.9091 & 1 & 24 & $6.379 \pm 1.036$ \\
    5.1855 & 2 & 41 & $0.518 \pm 0.941$ \\
    5.0184 & 2 & 28 & $0.778 \pm 0.932$ \\
    5.0065 & 2 & 27 & $0.515 \pm 0.761$ \\
    4.9947 & 2 & 26 & $1.015 \pm 1.042$ \\
    4.9831 & 2 & 25 & $0.768 \pm 1.054$ \\
    4.9716 & 2 & 24 & $1.193 \pm 1.122$ \\
    4.9490 & 2 & 22 & $0.918 \pm 0.871$ \\
    4.9379 & 2 & 21 & $0.777 \pm 0.892$ \\
    4.9269 & 2 & 20 & $1.373 \pm 1.021$ \\
    4.9161 & 2 & 19 & $1.211 \pm 0.827$ \\
    4.9054 & 2 & 18 & $1.194 \pm 0.912$ \\
        \hline
    \end{tabular}
    \label{tab:CO_fluxes}
\end{table}

%-------------------------------------------------------------------

\section{$\chi^2$ maps}
\label{app:chi2_maps}

\begin{figure*}[h]
\begin{tabular}{cc}
  \includegraphics[width=0.43\textwidth]{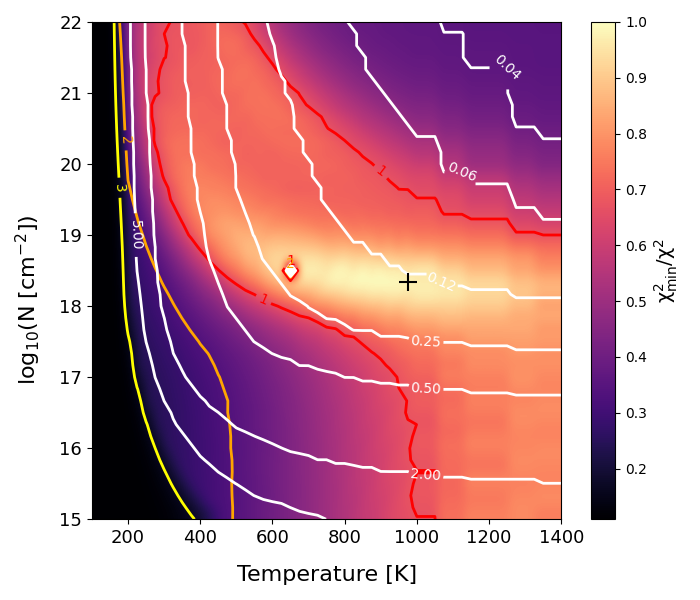} &   \includegraphics[width=0.43\textwidth]{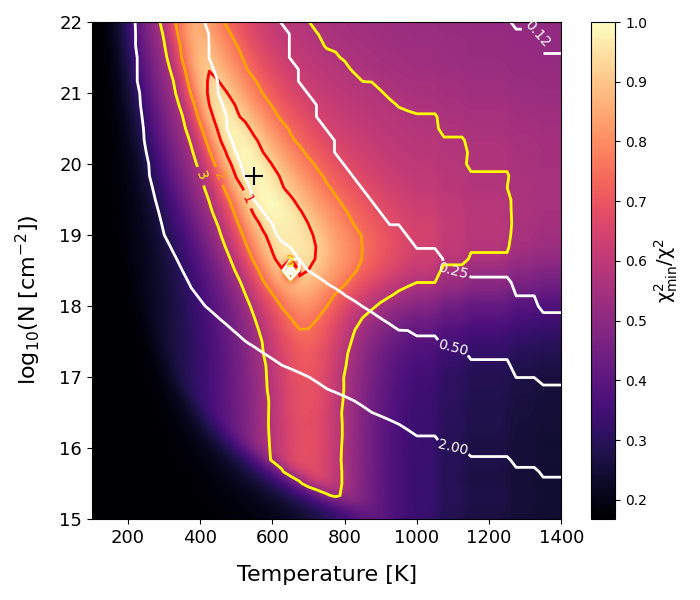} \\
(a) \H2O from 6\,\micron\ to 8\,\micron & (b) \H2O from 13\,\micron\ to 17\,\micron \\[6pt]
 \includegraphics[width=0.43\textwidth]{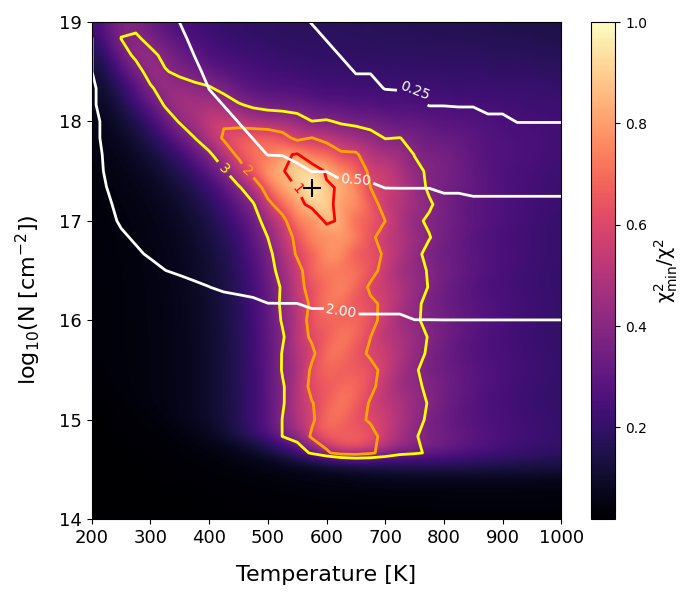} &   \includegraphics[width=0.43\textwidth]{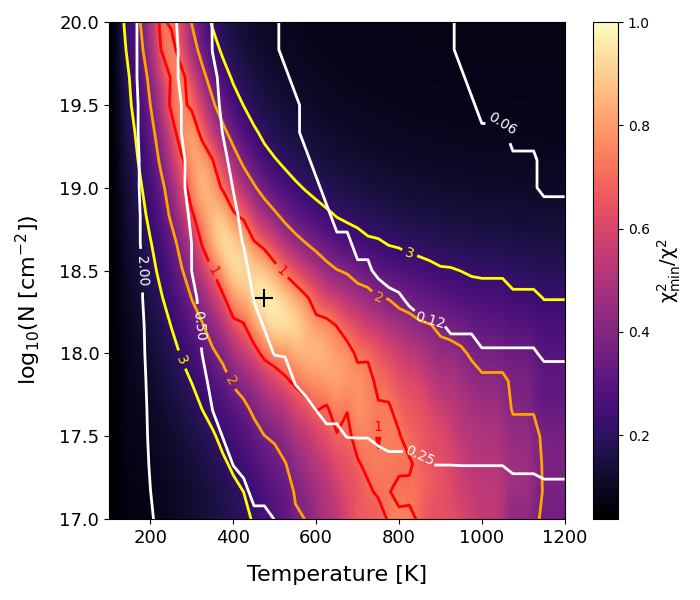} \\
(c) HCN & (d) \C2H2 \\[6pt]
  \includegraphics[width=0.43\textwidth]{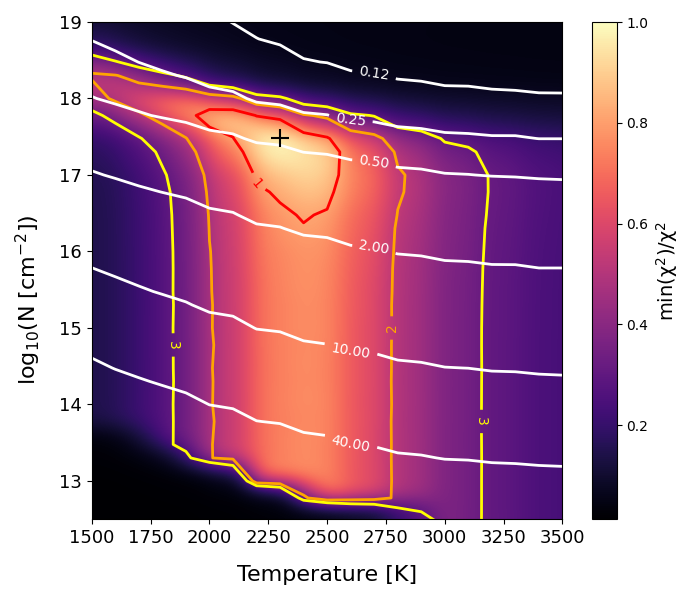} &   
  \includegraphics[width=0.43\textwidth]{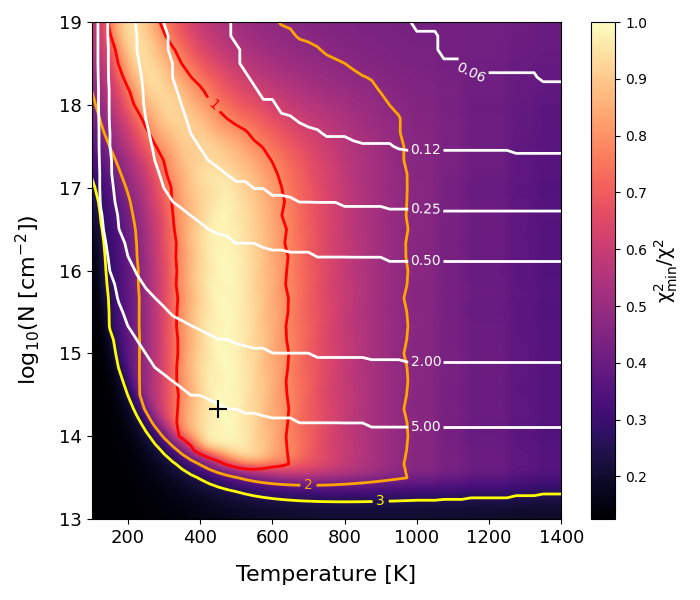} \\
(e) CO & (f) \CO2
\end{tabular}
\caption{$\chi^2$ maps resulting from the fitting of LTE slab models to the MIRI MRS spectrum (a -- f). The colors indicate the $\chi^2_{\rm min}/\chi^2$ value, being 1 the best-fit model. The white contours show the emitting radius in au and the red, orange and yellow contours show the 1, 2, and 3$\sigma$ confidence intervals. The location of the best-fit models in the parameter space is shown by a black cross.}
\end{figure*}

%-------------------------------------------------------------------
\clearpage
\section{Comparison of line luminosities with other samples}

\begin{figure*}[ht]
\centering
\includegraphics[width=\hsize]{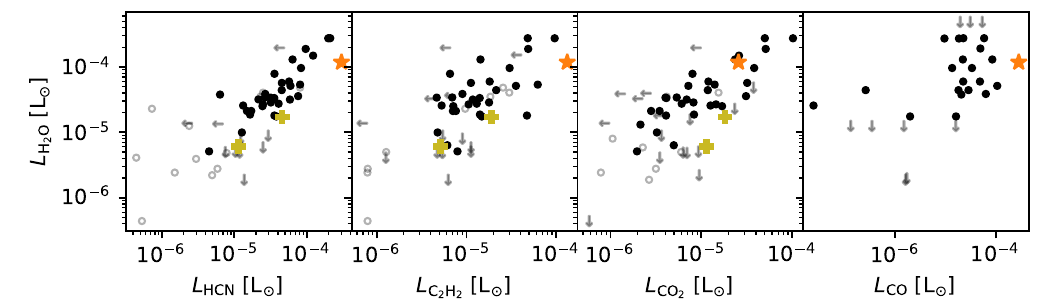}
  \caption{Comparison between the gas line luminosities of XUE\,1 and the \textit{Spitzer} sample studied by \citet{2020ApJ...903..124B} and the CO emission from \citet{2022AJ....163..174B} (black dots). The y-axis shows the 17\,\micron\ \H2O\ luminosity. From left to right the x-axis show the \CO2, HCN, \C2H2 and CO luminosities. The upper limits are indicated with grey arrows and the empty circles show the sources with upper limits both in the x and in the y-axis. The blue crosses show the location of the MIRI spectra published to date; GW\,Lup \citep{2023ApJ...947L...6G} and EX\,Lup \citep{2023arXiv230108770K}.The luminosities measured for the JWST spectrum of XUE\,1 are shown with the orange stars.}
     \label{fig:Banzatti20}
\end{figure*}

\bibliography{references}{}
\bibliographystyle{aasjournal}

\end{document}